\documentclass[
    ,final            
  ]
  {aipproc}

\layoutstyle{6x9}

\begin{document}

\title{
Nuclear isomers: structures and applications}

\classification{21.10.-k, 21.60.Cs, 26.30.+k}

\keywords{Nuclear isomers, rp-process, projected shell model,
multi-mass-zone X-ray burst model}

\author{Yang Sun}{
address={Department of Physics and Joint Institute for Nuclear
Astrophysics, University of Notre Dame, Notre Dame, Indiana 46545,
USA} }
\author{Michael Wiescher}{
address={Department of Physics and Joint Institute for Nuclear
Astrophysics, University of Notre Dame, Notre Dame, Indiana 46545,
USA} }
\author{Ani Aprahamian}{
address={Department of Physics and Joint Institute for Nuclear
Astrophysics, University of Notre Dame, Notre Dame, Indiana 46545,
USA} }
\author{Jacob Fisker}{
address={Department of Physics and Joint Institute for
Nuclear Astrophysics, University of Notre Dame, Notre Dame,
Indiana 46545, USA} }

\begin{abstract}
Isomeric states in the nuclei along the rapid proton capture
process path are studied by the projected shell model. Emphasis is
given to two waiting point nuclei $^{68}$Se and $^{72}$Kr that are
characterized by shape coexistence. Energy surface calculations
indicate that the ground state of these nuclei corresponds to an
oblate-deformed minimum, while the lowest state at the
prolate-deformed minimum can be considered as a shape isomer. Due
to occupation of the orbitals with large $K$-components, states
built upon two-quasiparticle excitations at the oblate-deformed
minimum may form high $K$-isomers. The impact of the isomer states
on isotopic abundance in X-ray bursts is studied in a
multi-mass-zone X-ray burst model by assuming an upper-lower limit
approach.
\end{abstract}

\maketitle


\section{Introduction}

It has been suggested that in X-ray binaries, nuclei are
synthesized via the rapid proton capture process (rp-process)
\cite{VanWorm,rpReport}, a sequence of proton captures and $\beta$
decays responsible for the burning of hydrogen into heavier
elements. Recent reaction network calculations \cite{rp} have
shown that the rp-process can extend up to the heavy Sn-Te mass
region. The rp-process proceeds through an exotic mass region with
$N\approx Z$, where the nuclei exhibit unusual structure
properties. Since the detailed reaction rates depend on the
nuclear structure, information on the low-lying levels of relevant
nuclei is thus valuable for the isotopic abundance study.

Depending on the shell filling, some nuclei along the rp-process
path can have excited metastable states, or isomers \cite{Nature},
by analogy with chemical isomers. Of particular interest are two
kinds of isomers, as illustrated in Fig. 1. It is difficult for an
isomeric state either to change its shape to match the states to
which it is decaying, or to change its spin orientation relative
to an axis of symmetry. Therefore, isomer half-lives can be very
long. If such states exist in nuclei along the rp-process path,
the astrophysical significance could be that the proton-capture on
long-lived isomers may increase the reaction flow, thus reducing
the timescale for the rp-process nucleosynthesis during the
cooling phase.

Coexistence of two or more stable shapes in a nucleus at
comparable excitation energies has been known in nuclei with A
$\approx 70 - 80$. The expected nuclear shapes include, among
others, prolate and oblate deformations. In an even-even nucleus,
the lowest state with a prolate or an oblate shape has quantum
numbers $K^\pi$ = $0^+$. An excited 0$^+$ state may decay to the
ground 0$^+$ state via an electric monopole (E0) transition. For
lower excitation energies, the E0 transition is usually slow, and
thus the excited 0$^+$ state becomes a ``shape isomer". There are
also excited states based on two-quasiparticle (qp) excitation. If
the two quasiparticles occupy the orbitals having large $K$ (where
$K$ is the quantum number representing the projection of the total
nuclear spin along the symmetry axis), the decay path to lower
energy states having a small or zero $K$ requires a large change
in $K$ quantum number, and therefore the emission of radiation
with high multipolarity is required to match the change. Such
emissions are usually strongly hindered, and thus the excited
high-$K$ state becomes a ``$K$-isomer" (see the schematic
illustration in Fig. 1).

\begin{figure}
\includegraphics*[angle=0,width=24pc]{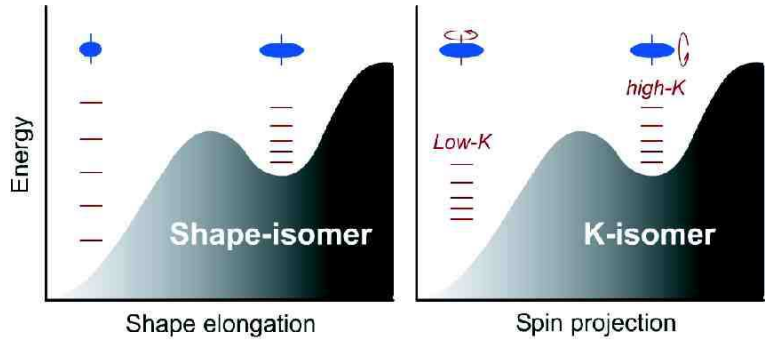}
\caption{Schematic illustration for nuclear excitation as
functions of various nuclear variables. The secondary energy
minima are responsible for the different kinds of isomers.}
\end{figure}

\section{Shape-isomers and K-isomers in $^{68}$Se and $^{72}$Kr}

\begin{figure}
\includegraphics*[angle=0,width=24pc]{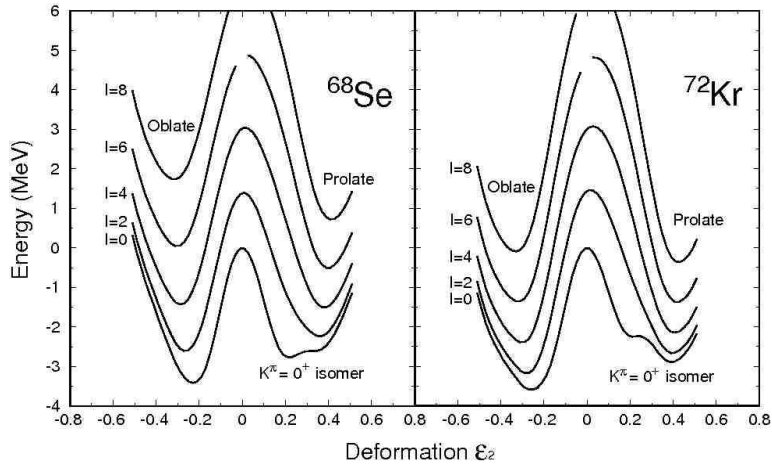}
\caption{Energy surfaces for various spin states in $^{68}$Se and
$^{72}$Kr as functions of deformation variable $\varepsilon_2$.}
\end{figure}

Calculations on the structure are performed by the projected shell
model \cite{psm}. Fig. 2 shows calculated total energies as a
function of deformation variable $\varepsilon_2$ for different
spin states in $^{68}$Se and $^{72}$Kr. The configuration space
and the interaction strengths in the Hamiltonian are the same as
those employed in the previous calculations for the same mass
region \cite{Sun04}. Under these calculation conditions, it is
found that in both nuclei, the ground state takes an oblate shape
with $\varepsilon_2\approx -0.25$. As spin increases, the oblate
minimum moves gradually to $\varepsilon_2\approx -0.3$. Another
local minimum with a prolate shape ($\varepsilon_2\approx 0.4$) is
found to be 1.1 MeV ($^{68}$Se) and 0.7 MeV ($^{72}$Kr) high in
excitation. Bouchez {\it et al.} \cite{BouchezKr72} observed the
671 keV shape-isomer in $^{72}$Kr with half-life $\tau = 38\pm 3$
ns. The one in $^{68}$Se is our prediction, awaiting experimental
confirmation. Similar isomer states have also been calculated by
Kaneko {\it et al.} \cite{Kaneko04}.

Most nuclei near the $N=Z$ line with $A\sim 70 - 80$ are
well-deformed. At the deformed potential minimum, the high-$j$
$g_{9/2}$ orbit intrudes into the $pf$-shell. With an oblate
shape, one finds the largest $K$ components ($K={7\over 2}$ and
$9\over 2$) of this $j$-orbit near the Fermi levels of $^{68}$Se
and $^{72}$Kr. Thus, a 2-qp state can have $K={7\over 2}+{9\over
2}=8$, and a 4-qp state $K=16$ which is built from a neutron 2-qp
and a proton 2-qp state. If $K$ is approximately a conserved
quantum number, the $K$ value in these 2- and 4-qp states is much
larger than that of the ground state band ($K=0$). Once having
been populated, this makes it rather difficult for such 2- or 4-qp
states to decay back to the ground state.

In Figs. 3, we present the energy levels calculated by the
projected shell model, and compare them with available
experimental data \cite{FischerPRL}. For $^{72}$Kr, with the newly
confirmed $0^+$ isomer \cite{BouchezKr72} which should be the
bandhead of the prolate band, the rotational band at the prolate
minimum is now known. However, there have been no experimental
data to compare with the predicted oblate band. In contrast, an
oblate band in $^{68}$Se was observed and a prolate one was also
established \cite{FischerPRL}, except for the missing bandhead
which we predict as a shape isomer. For both nuclei, we predict
low-lying high-$K$ isomers, indicated by bold lines. In
particular, the spin-16 states are so low in excitation (much
lower than the spin-16 state in the ground band) that one may
consider them as a spin trap \cite{Nature}.

\begin{figure}
\includegraphics*[angle=0,width=31pc]{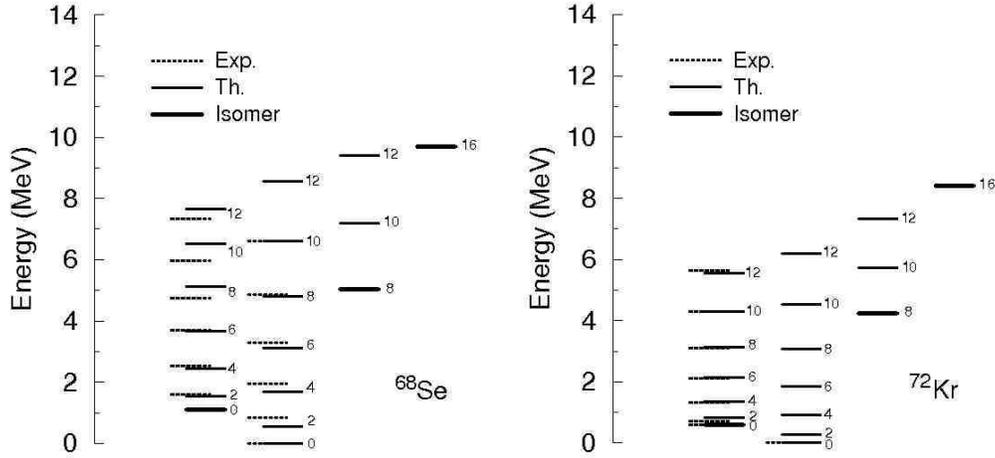}
\caption{Calculated energy levels for $^{68}$Se and $^{72}$Kr by
the projected shell model. The theoretical results are compared
with available data.}
\end{figure}

\section{Impact on isotopic abundance in X-ray bursts}

The recent observation of a low energy 0$^+$ shape isomer in
$^{72}$Kr \cite{BouchezKr72} has opened new possibilities for the
rp-process reaction path. A similar shape isomer has been
predicted for $^{68}$Se in this paper. Since the ground states of
$^{73}$Rb and $^{69}$Br are bound with respect to these isomers,
proton capture on these isomers may lead to additional strong
feeding of the $^{73}$Rb($p,\gamma$)$^{74}$Sr and
$^{69}$Br($p,\gamma$)$^{70}$Kr reactions. However, whether these
branches have any significance depends on the associated nuclear
structure parameters, such as

\begin{itemize}
\item{how strong is the feeding of the isomer states?}

\item{what is the lifetime of the isomer with respect to
$\gamma$-decay and also to $\beta$-decay?}

\item{what are the lifetimes of the proton unbound $^{69}$Br and
$^{73}$Rb isotopes in comparison to the proton capture on these
states?}
\end{itemize}

Two processes can be envisioned to populate the isomeric states in
appreciable abundance, through thermal excitation of the ground
state at high temperatures, or through proton capture induced
$\gamma$-feeding. Thermal excitation is very efficient for feeding
levels at low excitation energy since the population probability
scales with $e^{-E_{is}/kT}$. Contributions of low energy states
(E$_x\le$Q) are negligible since proton capture on those states is
balanced by inverse proton decay \cite{rpReport}. This is not the
case for proton capture on the isomeric states. The peak
temperature in the here used X-ray burst model is around 1.1 GK,
the isomer states in $^{68}$Se at 1.1 MeV and in $^{72}$Kr at 0.67
MeV are therefore only very weakly populated with $\le 0.02\%$ and
$\le 0.5\%$, respectively. Feeding through
$^{67}$As(p,$\gamma$)$^{68}$Se$^*$ (Q$\approx$3.19 MeV) and
$^{71}$Br(p,$\gamma$)$^{72}$Kr$^*$ (Q$\approx$4.1 MeV) is a more
likely population mechanism. A quantitative prediction of the
feeding probability requires a more detailed study of the
$\gamma$-decay pattern of low spin (J$\le$3) states above the
proton threshold in $^{68}$Se and $^{72}$Kr, respectively.

The lifetime of the isomeric states must be sufficiently long to
allow proton capture to take place. No information is available
about the lifetime of the $^{68}$Se$^*$ isomer while the 55 ns
lifetime of the isomer in $^{72}$Kr is rather short
\cite{BouchezKr72}. Based on Hauser Feshbach estimates
\cite{rpReport} the lifetime against proton capture is in the
range of $\approx$100 ns to 10 $\mu$s depending on the density in
the environment. Considering the uncertainties in the present
estimates a fair fraction may be leaking out of the $^{68}$Se,
$^{72}$Kr equilibrium abundances towards higher masses.

This however also depends on the actual proton decay lifetimes of
$^{69}$Br and $^{73}$Rb. Based on model dependent fragmentation
cross section predictions for these isotopes lifetimes have been
estimated to be less than 24 ns and 30 ns respectively
\cite{pfaff}. Again, within the present systematic uncertainties
this is in the possible lifetime range of proton capture processes
in high density environments .

\begin{figure}
\includegraphics*[angle=0,width=27pc]{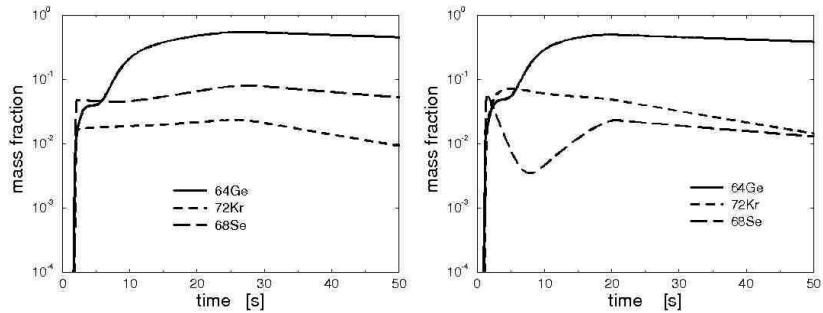}
\caption{Mass fractions in the X-ray burst model with two extreme
cases. Left: no isomer contribution. Right: full isomer
contribution.}
\end{figure}

While it is likely that equilibrium is ensued between all these
configurations within the presently given experimental limits a
considerable flow towards higher masses through the isomer branch
cannot be excluded. Fig. 4 shows the comparison between the two
extreme possibilities for the reaction sequence calculated in the
framework of a multi-mass-zone X-ray burst model \cite{fisker}.
The left-hand figure shows the mass fractions of $^{64}$Ge,
$^{68}$Se, and $^{72}$Kr as a function of time neglecting any
possible isomer contribution to the flow. The right-hand figure
shows the results from the same model assuming full reaction flow
through the isomeric states in $^{68}$Se and $^{72}$Kr rather than
through the respective ground states. The K-isomers predicted in
this paper have not been considered. The main differences in
$^{68}$Se and $^{72}$Kr mass fractions are due to rapid initial
depletion in the early cooling phase of the burst. This initial
decline is compensated subsequently by decay feeding from the long
lived $^{64}$Ge abundance. The results of our model calculations
shown in Fig. 4 are based on upper and lower limit assumptions
about the role of the shape isomer states. The possible impact on
the general nucleosynthesis of $^{68}$Se and $^{72}$Kr turns out
to be relatively modest. These assumptions are grossly simplified.
Improved calculations would require better nuclear structure data
to identify more stringent limits on the associated reaction and
decay rate predictions.

We are just beginning to look at the impact that isomers may have
on various nucleosynthesis processes such as the rp-process. We
look for cases in which an isomer of sufficiently long lifetime
(probably longer than microseconds) can change the paths of
reactions taking place and lead to a different set of elemental
abundances. This aspect of nuclear isomers is very much in its
infancy \cite{Ani05}.

\begin{theacknowledgments}
This work is partly supported by NSF under contract PHY-0140324.
\end{theacknowledgments}

\end{document}